\documentclass[preprint2]{aastex}

\slugcomment{Submitted to PASP}

\shorttitle{Overlapping Galaxy Pairs}
\shortauthors{Keel et al.}

\begin{document}


\title{Galaxy Zoo: A Catalog of Overlapping Galaxy Pairs for Dust Studies}


\author {William C. Keel\altaffilmark{1,2} and Anna M. Manning\altaffilmark{1,3}}
\affil{Department of Physics and Astronomy, University of Alabama,
Box 870324, Tuscaloosa, AL 35487}
\email{wkeel@ua.edu}

\author{Benne W. Holwerda}
\affil{ESA-ESTEC, Keplerlaan 1, 2201 AZ Noordwijk, The Netherlands}

\author{Massimo Mezzoprete}
\affil{Rome, Italy}

\author{Chris J. Lintott}
\affil{Astrophysics, Oxford University; and Adler Planetarium, 1300 S. Lakeshore Drive, Chicago, IL 60605}

\author{Kevin Schawinski\altaffilmark{4}}
\affil{Department of Physics, Yale University, New Haven, CT 06511 USA; and Yale Center for Astronomy and Astrophysics, Yale University, P.O.Box 208121, New Haven, CT 06520 and Institute for Astronomy, Department of Physics, ETH Zurich, Wolfgang-Pauli-Strasse 16, CH-8093 Zurich, Switzerland}

\author{Pamela Gay}
\affil{Center for STEM REO, Campus Box 2224, Southern Illinois University Edwardsville, Edwardsville, Illinois 62026}

\and 

\author{Karen L. Masters}
\affil{Institute of Cosmology and Gravitation, University of Portsmouth,
Dennis Sciama Building, Burnaby Road, Portsmouth, PO1 3FX, UK}

\altaffiltext{1}{Visiting Astronomer, WIYN Observatory, Kitt Peak National Observatory.
KPNO is operated by AURA, Inc. under contract to the National Science
Foundation}
\altaffiltext{2} {SARA Observatory}
\altaffiltext{3} {Now at Naval Oceanographic Office, Stennis Space Center, MS}
\altaffiltext{4} {Einstein Fellow}



\begin{abstract}
Analysis of galaxies with overlapping images offers a direct way to probe the
distribution of dust extinction and its effects on the background light. We present
a catalog of 1990 such galaxy pairs selected from the Sloan
Digital Sky Survey (SDSS) by volunteers of the Galaxy Zoo project.
We highlight subsamples which are particularly useful for retrieving such
properties of the dust distribution as UV extinction, the extent perpendicular to the disk plane, 
and extinction in 
the inner parts of disks. The sample spans wide ranges of morphology and surface brightness, 
opening up the possibility of using this technique to address systematic changes in
dust extinction or distribution with galaxy type. This sample will form the basis for forthcoming
work on the ranges of dust distributions in local disk galaxies, both for their astrophysical 
implications and as the low-redshift part of a study of the evolution of dust properties. 
Separate lists and figures show deep overlaps, where the inner regions of the foreground galaxy are
backlit, and the relatively small number of previously-known overlapping pairs
outside the SDSS DR7 sky coverage.
\end{abstract}




\section{Introduction}

Dust composes only a small fraction of the mass of the interstellar medium in
galaxies, but it plays key roles in both their development and observed properties. Radiative cooling by dust 
facilitates the collapse of massive clouds during episodes of star formation, and
grains remain as a key constituent of protoplanetary disks around the nascent stars.
At the other end of the starsÕ lives, grains condense in the cooling and enriched gaseous
envelopes of supergiants, planetary nebulae, and the ejecta of novae and
supernovae. On grander scales, dust influences 
the evolution of galaxies even as its absorption complicates our ability to
interpret the observed properties of galaxies. Correction for dust
extinction within galaxies is important for the use of SN Ia as cosmological probes,
especially if the extinction curve varies among galaxies. It also affects distance
measures via the Tully-Fisher relation, through the required corrections for internal
extinction as a function of inclination to our line of sight. If the extinction is
high enough toward the cores of luminous spirals, many straightforward estimates
of stellar masses and distributions come into question \citep{Driver}. We present 
an extensive catalog of superimposed galaxy pairs, which are useful
for measuring dust extinction and allow a much wider study than previously available. 

For decades, the standard view on dust effects in galaxies derived from variants of the
\citet{Holmberg} approach, based on surface brightnesses of disks viewed at
different inclinations to the line of sight. Optically thin disks will increase in surface
brightness when viewed closer to edge-on, while in the optically thick limit, the
surface brightness will remain constant. Holmberg's data indicated a mild inclination
dependence and correspondingly modest internal extinction. This comforting
conclusion was eventually challenged by \citet{Disney} on
theoretical grounds; models with very dusty disks could reproduce the data if the
disk scale height of the dust is much less than that of the stars. Around the same
time, \citet{Valentijn} claimed that extensive analysis of surface photometry of galaxies from the 
ESO-LV sample suggested nearly constant surface brightness with disk
inclination, and much higher internal opacity than previously assumed. These
results spurred a rebirth in work on this question, leading to a wide range of studies
converging on a general picture of high opacity in spiral arms, resonance rings, and
the centers of disks, with diffuse dust fading to very small extinction at the edges of
the optical disks \citep{Thronson}. Still, it became clear that substantial variations
exist among and within galaxies. Use of the Galaxy Zoo morphological classifications
with color data suggested further complications, such as very luminous disks being
comparatively dust-deficient \citep{Masters}.

All these factors can be exacerbated for galaxies at high
redshift. Many have high rates of star formation, suggesting a rich and massive
interstellar medium, and are observed most often in the
emitted ultraviolet, where dust effects are more important, compared to the
emitted optical range where our basis for comparison is more extensive. Different ways of correcting for
internal extinction give widely varying results, contributing to the disparate
conclusions obtained by several groups on the history of cosmic star formation,
even when starting from the same Hubble data in the deep fields (e.g., \citealt{Madau},
\citealt{Thompson}).

The dust content and distribution in galaxies may be approached
in several ways.
Together with models of radiative transfer and constraints from the global energy budget,
fits to the spectral energy distributions (SEDs) of galaxies have been used to estimate the
typical extinction at various wavelengths and total dust masses. Models must assume
the relative distributions of stars and dust, as set out by, for example, \citet{WTC};
dust mixed with the stars gives much less overall extinction than a simple foreground screen.
This must be taken into account in deriving extinction laws from comparison of
objects' spectra, as in the effective extinction law derived by \citet{Calzetti}. SED fitting
of edge-on spirals consistently underestimates the face-on extinction and amount of
UV radiation which is reprocessed
 (\citealt{Baes2003}, \citealt{Baes2010}, \citealt{Kuchinski}),
 so improved measurements of these quantities are important in refining models of this kind.

Far-infrared emission provides very direct mapping of the dust content of
galaxies, modulated by the typical grain temperature. Very cold grains remain
difficult to detect, and large fractions of the total dust mass can in principle be
found only by careful spectral decomposition of the submillimetre emission, in view of the
dominant contribution of the warmest grains. Far-IR data
have hinted that some spirals have very extensive dust distributions, at least 
through a larger disk scale length for dust than for starlight 
(\citealt{Davies1999}, \citealt{Sun}). Aside from 
the detectability of dust being strongly weighted to warmer grain populations, 
far-IR emission studies still suffer from having spatial resolutions much poorer than is
available using optical techniques.

{\it Herschel} observations have contributed considerably to understanding the
radial and temperature distributions of galactic dust, and the total mass in grains.
With adequate spatial resolution, the mix of grain temperatures should be coupled to
location in the galaxy through the radiation field of starlight.
{\it Herschel} traces dust at 350 $\mu$m out to 0.8--1.3 times the diameter $D_{25}$ of
the de Vaucouleurs isophote in luminous spirals (\citealt{Bendo}, \citealt{Pohlen}).
While the gas/dust ratio increases with radius in Virgo spirals, as might be expected from metallicity
gradients (\citealt{Smith}, \citealt{Pohlen}), the dust shows a cutoff
matching that of H I in stripped galaxies \citep{Cortese}, testifying to a dynamical
link between gas and dust. Typical {\it Herschel} observations reach dust column densities
close to 0.05 $M_\odot$ pc$^{-2}$ at $T=19-22$ K (e.g. \citealt{Smith}). Even with these dramatic improvements in sensitivity and angular resolution, 
degeneracies remain in interpreting the grain populations in galaxies. These arise largely
because changes in the emissivity parameter $\beta$ can mimic changes in the relative contributions of
warm and cold dust across the FIR and submm regimes. The value of $\beta$ expresses
departures from the blackbody intensity $B_\lambda (T)$ for an emitter which is small compared to the 
relevant wavelengths, of the form $I \propto \lambda^{- \beta} B_\lambda (T)$. Changes in the distribution of
grain size will affect $\beta$, although it remains unclear whether this would be measurable when
averaged across whole galaxies (\citealt{Alton}, \citealt{James}).

Absorption studies have the high angular resolution of optical or ultraviolet imaging,
but are limited to regions with background sources that are adequate in solid-angle coverage 
and our understanding of their properties. When the background light arises within the
 galaxy under study, other issues arise in our knowledge of the relative distribution of stars and dust,
 and the role of scattering; these are approached in such ways as modeling the $z$-distribution of disk stars of
 various kinds to interpret multicolor absorption measurements \citep{Elmegreen1980}.
 In comparison, use of a more distant galaxy as the backlighting source gives several advantages, at the
 expense of dramatically reducing the range of galaxies available for analysis. The spatial resolution
 is limited only by the telescope's image quality, and we need know nothing about the internal
 stellar structure of the foreground galaxy. Scattering corrections are negligible 
 once the two galaxies have a line-of-sight separation only a few times their diameters, satisfied
 if their redshift difference indicates that the galaxies are far apart along the line of sight. High-quality 
 extinction measurements can be obtained even in the outskirts of a galaxy disk. 
 
Results of both these approaches
 to absorption will have a clumping dependence; internal structure in absorption can make
 small clumps harder to detect if they are below the spatial resolution of the data, and the
 derived reddening law will generally be flatter (grayer) than the intrinsic form given
 by the grain properties, due to mixing of regions with different transparency within a single
 resolution element. The dust clumping also enters into the distinction between dust mean column
 density as derived from modeling FIR measurements and the optical extinction we measure here.
 
Application of the overlap technique has improved with data quality. \citet{Keel83} showed a first application to NGC 3314 and the foreground system in NGC 1275, with vidicon imagery of limited signal-to-noise ratio. This approach was put on a genuinely quantitative footing with 
analyses of CCD imagery by \citet{Andredakis}, \citet{WK1992},  \citet{Berlind}, \citet{WKC2000} and \citet{Domingue99} (D1999) , in the last case comparing optical extinction with far-IR and submillimeter
data to compare dust masses estimated from emission and absorption. 
Broadly, these studies show that extinction may be high ($A_B > 1$) within spiral arms at a wide range of galactocentric radii, with interarm
extinction smaller and declining outward in a roughly exponential manner. Extinction in the inner kpc has been measured only in NGC 3314a, where
it reaches $A_B > 5$. One resonance-ring spiral was observed, in which the ring opacity is much larger than that found on either side of it. 
Extending
the analysis to the resolution of tens of parsecs enabled by the {\it Hubble Space Telescope} (\citealt{KW2001a}, \citealt{KW2001b}, \citealt{Elmegreen}, \citealt{Holwerda2009}) showed that the effective reddening law depends significantly on linear resolution. For a clumpy distribution of dust, this is unavoidable, since the weighting of
regions in transmission becomes wavelength-dependent \citep{Fischera2003}. The values measured in the outer disks of several spirals at HST resolution 
approach the local Milky Way mean, suggesting that the extinction contrast of dust structure on these
scales is modest enough that we may plausibly be approaching the true grain reddening behavior. These
studies were all limited by the small number of suitable and nearby galaxy pairs known, so that 
the behavior with galaxy type and its variance could not be 
explored. In particular, only the most symmetric spirals - grand-design and strongly ringed systems - could be
measured when the sample size was too small to average over unmodelled structure in individual galaxy disks.

The ideal pair for mapping dust would consist of a face-on, fairly symmetric spiral, seen nearly half backlit by a smooth elliptical or S0 system, of which half
is seen free of any foreground extinction. In such a system, there is a large region in which each galaxy is seen essentially by itself so that a good model of each galaxy can be produced, and a large region over which dust can be mapped. If $B$ is the background intensity and $F$ the foreground intensity, modelled point-by-point, the optical depth $\tau$ may be estimated from the observed intensity $I$ using
$$ e^{- \tau} = {{I - F} \over {B}}$$
Various compromises may have to be allowed for some uses; for example, only in spiral/spiral overlaps is there enough flux to measure extinction into the ultraviolet, although the point-by-point errors are unavoidably large due to the rich structure of spirals which is most pronounced at such short
wavelengths. If they are seen through particular parts of a spiral, even background galaxies of comparatively small angular size may be useful (if seen at the disk edge or across a spiral arm, for example); selection effects dictate care in analysis because of the small objects that will not be seen through
relatively opaque regions \citep{Holwerda2007c}. Useful spectroscopic estimates of extinction can be made even when there are
significant departures from symmetry, using the relative amounts of light at each galaxy's redshift (\citealt{Domingue00}; D2000).

The ability to harvest large samples of overlapping
galaxies from surveys like the SDSS has reinvigorated the study of extinction using backlighting.
We present here a new, large listing of galaxy pairs suitable for such dust studies. Its production has 
relied on contributions from many of the volunteer participants in the Galaxy Zoo project \citep{Lintott2008}. We will use this sample to address a range of issues in galaxy dust content via this single technique. What are the systematics with Hubble type, central surface brightness, strength of bars or rings? How common are the kinds of extended dusty disks found by \cite{Holwerda2009}?  Is there a very dusty morphological disk sequence which doesn't stand out in color alone \citep{Thronson}? In addition to the study of dust extinction in a variety of environments, uses of the listing range from testing image-decomposition routines to correlation with supernova detections as a way of measuring reddening laws.

\section{Galaxy Selection}
The bulk of this catalog consists of pairs originally selected by volunteer members of the Galaxy Zoo project\footnote{http://www.galaxyzoo.org}.  As described by \citet{Lintott2008}, the main goals
of the project are reached by visual classification of galaxies from the SDSS via a web interface. This project operated outside the primary statistically-oriented framework 
of the Galaxy Zoo project, as so-called forum science. After noticing that some of the
pairs being shown and discussed on the project forum\footnote{http://www.galaxyzooforum.org} were likely non-interacting, overlapping pairs, WCK posted a specific request for such systems on the forum. Users responded enthusiastically, posting images and identifications for large numbers of
candidates\footnote{The dedicated discussion thread for these pairs is http://www.galaxyzooforum.org/index.php?topic=6732.0}. We inspected these candidate pairs (and objects appearing on other discussion threads of the forum, a total of $\approx 7000$ suggested pairs) for overlap, utility for dust study, and evidence of interaction or
other asymmetry, with surviving candidates going into our working list. 
Galaxy Zoo visual detections are now by far the dominant source for low-redshift overlapping systems. Prior to surveys from the SDSS images
(\citealt{HKB2007} and this work), no more
than 25 such pairs were mentioned in the literature (Appendix A). 

For completeness, we checked for pairs already selected  spectroscopically  from the SDSS \citep{HKB2007}. Modifying the criteria used for the 
gravitational-lens search by \citet{Bolton2004}, that project selected objects where the SDSS spectra showed an emission-line redshift lower than the cross-correlation (absorption-line) value. This target list was winnowed visually for appropriate geometry (images actually overlapping) to
give a final set of 83 pairs. Of these, 34 are not useful morphologically for our sample here, and 9 were already selected by the Galaxy Zoo project. The remainder
were objects in which two galaxies are not apparent on inspection of the SDSS images, such as having a low-luminosity foreground object closely aligned with a luminous background galaxy (so our image analysis is not appropriate, at least with
ground-based data).

Some of the newly-recognized systems are large enough and bright enough to appear in earlier
surveys (such as the Palomar Sky Survey), but the depth of the SDSS data and the helpful generation
of color-composite images with high-dynamic-range mapping \citep{lupton} made it easy to
recognize features that were overexposed on most digital representations of the Palomar photographs.
The color information was especially valuable for finding reddened galaxies behind the disks of
spirals, which may be prominent only in the $i$ and $z$ bands.

As an additional check for automated selection of overlapping pairs, Galaxy 
Zoo volunteer Lionel P\"offel contributed the results of an SQL query to identify
pairs of objects closer than the sum of their listed Petrosian radii, with
redshift difference $\Delta z > 0.01$. With hindsight, this approach was
rendered less effective by blending issues; for strongly overlapping galaxy images, the SDSS photometric pipeline often extracts only the nucleus of one galaxy, assigning a Petrosian radius too small by factors 4--5. A 
converse issue is that stars can be misclassified as galaxies when superimposed on the outskirts of a large galaxy image. Indeed, of
120 candidate pairs from this selection, 24 involve a foreground star. Even after removing these
false alarms, this query added 28 new pairs to the catalog, and recovered 14 otherwise selected (plus 49 objects which had been considered from forum postings and
rejected for reasons of symmetry, low surface brightness in the overlapping
regions, or inappropriate orientations). This comparison testifies to the utility of visual selection by the Galaxy Zoo volunteers in
generating a candidate sample for this project.

Not all galaxies in our catalog are in the SDSS database. A few were otherwise known from earlier results and lie outside its footprint (Appendix A). 
Even within the imaging coverage, some have at least one member not identified as a separate photometric object. 

Of 1990 pairs comprising our final catalog, 1713 (86\%) have one or more known redshifts measured in the SDSS or tabulated by NED\footnote{http://ned.ipac.caltech.edu/}. The NED
additions are for bright galaxies or companions to galaxies with SDSS spectra, where there is no SDSS spectrum because of sampling
or fiber-collision limitations. 
We adopt a redshift-difference criterion to flat the subset of pairs which cannot be physically interacting, driven by the range of $\Delta z$ seen in
tidally-interacting galaxy pairs and including high-velocity encounters in clusters. We adopt $\Delta z > 0.008$ ($
\Delta v > 2400$ km s$^{-1}$, a very conservative value which would distinguish non-associated galaxies even in the presence
of significant redshift errors. Of 494 pairs with both redshifts known, 218 satisfy this criterion, ruling out gravitational interaction
with each other. For this subset, scattering effects are guaranteed to be negligible in measurements of obscuration \citep{WKC2000}, and there 
will be no asymmetries driven by interaction between the pair members. We rejected candidate pairs which have these
large redshift differences, but in the wrong sense
to be useful for dust studies (such as an elliptical galaxy in the foreground of a spiral).

\begin{figure} 
\includegraphics[width=80.mm,angle=0]{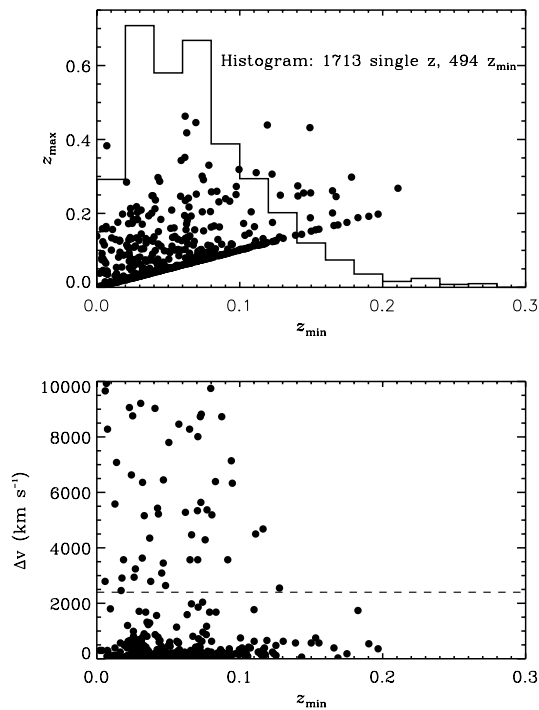} 
\caption{Distribution of known redshifts of catalogued pair members. Points indicate lower and higher redshift values
$z_{min}$ and $z_{max}$ when both are known. In the upper panel, the superimposed histogram shows the
distribution of single redshift values and $z_{min}$ in bins of $\Delta z$=0.02, with values scaled
down by factor 500 to fit the coordinate axes. The tight grouping of points along the diagonal
marks pairs with close enough redshifts to be physically associated on group scales, although not necessarily
interacting. This is emphasized in the lower panel, which shows the individual pair differences in radial velocity $\Delta v$.
The dotted horizontal line marks our $\Delta v = 2400$ km s$^{-1}$ division; pairs above this are not
physically associated.} 
\label{fig-zdist} 
\end{figure}

The catalog properties
in distance and $\Delta z$ are illustrated in Fig. \ref{fig-zdist}, which shows a histogram of
the smaller redshift value $z_{min}$, when both are known, or the single known redshift, and
plots values when both $z_{min}$ and $z_{max}$ are known. For systems with one redshift known, the median value is $z=0.070$, with quartiles 0.041, 0.105. Combining single redshifts with foreground
values when both are known, the median is $z=0.064$ (quartiles 0.036, 0.097), while for background
systems when both values are known, the range rises as expected: median $z=0.077$ and quartiles 0.040, 0.127.

WCK screened pairs posted by Galaxy Zoo participants for suitability, in particular rejecting pairs with obvious
tidal interaction as seen on the SDSS color composite images. In a final round, WCK and AMM
independently re-examined candidates, to check for clearly interacting pairs which had survived the initial round.
At this stage, we rejected about 10\% of systems initially selected for the catalog. These were
pairs with obvious tidal tails or decentered nuclei, objects where it was not clear whether there 
are in fact two distinct galaxies rather than unusual substructure in a single one, and pairs in which
the redshift difference is both in the wrong sense to be useful for dust studies (spiral behind
elliptical) and
 large enough to clearly show that the galaxies are not in the same
group, so that the redshifts will indicate distance ordering. This ``discard" list is available in case some of these systems prove of use for related questions.
We required positive evidence of tidal distortion to reject a candidate pair; some interacting
systems with only weak distortion, or tidal features evident only at low surface brightness, will remain in
the sample.  

Our catalog of 1990 overlapping galaxy pairs is listed in Table \ref{tbl-catalog}. The entire listing is
given in the online edition; the printed version includes a subset to illustrate content and format. This
subset includes the initial lines of the RA-ordered list, all the newly-found objects shown in Fig. \ref{fig-montage1}, and 
a few additional objects to illustrate the combinations of data and identifications found. Data for previously known pairs
in Fig. \ref{fig-montage1} are listed in Table \ref{tbl-oldpairs}.
SDSS pairs are listed with the coordinate designation of the brighter member. Magnitudes are SDSS  {\tt modelmag\_r} values in the $r$ band. 
These must be treated with caution, if not suspicion, because pipeline separation of strongly
blended galaxy images often results in assigning much of the fainter galaxy's flux to the brighter one (section 2.2), and spot checks
also show some
pairs in which even a smooth symmetric galaxy is broken into multiple SDSS photometric objects.. A second magnitude with only
one significant figure past the decimal was estimated from the SDSS composite image, if the galaxy was not detected as an SDSS object (or as
multiple objects). A value of 30.00 was assigned if the images are so strongly blended that no estimate for the fainter galaxy was reasonable.
In some cases there are multiple background galaxies; the magnitude listed is for the brightest of these. We include redshifts where features for
both members were clearly detected in a single SDSS spectrum (indicated by {\it fg} and {\it bg} for fore/background).
The catalog in sortable web form, with additional PDF files with
finding charts and additional data on each pair as well as our ``reject" list, is available from http://data.zooniverse.org/overlaps.html.

\subsection{Pair categories}

We note some special categories defined by geometry or redshifts with a ``type"
indicator in Table \ref{tbl-catalog}, which 
mark subsets of the catalog particularly useful for various purposes. They
are denoted by mnemonic designations, which are summarized along 
with their frequency in the catalog in Table \ref{summary}. A selection of each is illustrated in Fig. \ref{fig-montage1}.

{\bf F}: spirals seen nearly face-on in front of an elliptical or S0 background system. These are closest
to the ideal for most kinds of extinction studies. For our purposes, these include spirals face-on 
enough that the extinction structure is dominated by arm/interarm variations, rather than
the extent of dust perpendicular to the disk. In practice this includes 
spirals with planes inclined as much as 60$^\circ$ to the plane of the sky.

{\bf Q}: the background galaxy is nearly edge-on and is projected nearly radial 
to the foreground galaxy, so the backlit area spans a large range in radius but a narrow
one azimuthally within the foreground
system. This means that symmetry requirements on the
foreground galaxy can be relaxed substantially, modeling it by interpolation across a narrow angular sector. 
Arp 198 (UGC 6073, VV 267; 
Fig. \ref{fig-montage1}) is a good example, in which the run of extinction with radius can
be retrieved even in the presence of a rather complex foreground spiral pattern. In the
most favorable cases, the SDSS images show evidence of extinction to indicate that the
edge-on galaxy is in the background; better imagery may show some of the fainter Q systems
to in fact have the edge-on galaxy in front.

$\Phi$: the spiral is seen essentially edge-on, at least partially backlit by a smooth galaxy. These
are useful for studying the structure of extinction perpendicular to the disk of the foreground galaxy,
and the radial extent of extinction to very sensitive levels.
The letter is selected to remind one of a thin disk with a round galaxy behind it.
	
{\bf X}: both galaxies are seen nearly edge-on, with their disks crossing as 
seen either near both nuclei or along one disk. These are relevant to
the occurrence of extended $z$-distributions of dust (as in $\Phi$ types), as well as
to questions of the distributions of angular momenta in pairs (with 
geometrically-based corrections for the likelihood of seeing the disks 
intersect when they have various actual angular differences). A handful of 
these (coded with R, as below) may in fact be polar-ring systems seen edge-on to both the central
galaxy and the ring.
	
{\bf SE}: spiral/elliptical superpositions that do not fall in one of the other geometric categories above.

{\bf S}: spiral/spiral overlaps. These have much richer background structure than S/E overlaps, but
are useful in probing extinction into the ultraviolet because the background source will remain
detectable at much shorter wavelengths. At least using averaging techniques, they can be used to compare
optical and UV extinction. With high-quaIity images, additional symmetry clues such as the 
direction of arm features can help separate foreground and background structure \citep{KW2001b}.

{\bf B}: the background galaxy has much smaller angular size than the foreground disk. Such
alignments are universal for small enough background systems (transitioning, when their statistics are large, into the ``synthetic field" technique
matching galaxy counts and colors to global means and taking into account the cosmic variance; 
\citealt{Holwerda2005a}, \citealt{Holwerda2005b}, \citealt{Holwerda2005c}, \citealt{Holwerda2007a}). 
As these papers demonstrate, considerable care is needed using
numerous small background galaxies to map extinction, because of bias in favour of the more transparent regions. Therefore, our listing includes only those where the line of sight
to the background galaxy is particularly useful as a dust probe -- edge of the optical disk, inside a resonance ring, or behind an outer spiral arm. In these cases even a local measurement of differential extinction can be useful. This category in particular employed subjective criteria as to utility.

{\bf E}: pairs containing only elliptical or S0 galaxies, as judged from the SDSS. Our search concentrated
on spirals, so coverage of these systems is much less complete. Many such pairs occur in clusters, but
the possibility of subtle tidal distortions makes use of pairs with large redshift differences more valuable in seeking even low-opacity diffuse dust components.
 
{\bf R}: we use this code to flag a few objects which might be either
near-central superpositions or polar rings, where better imagery or spectroscopic information would be needed to be sure. All of these, if overlaps, have
very small central impact parameters and also appear in our list of deep overlaps.

Nineteen pairs do not fall in any of these geometrically-defined categories, and are labelled simply ``misc" in the table.

$\Delta z$: these are selected to have a known redshift difference so
large that the two galaxies will not be interacting with each other. 
Starting with known pairs, we take $\Delta z = 0.008$ as the demarcation point. These pairs are the least likely to show tidal features or have 
tidally-induced asymmetries, and are so far apart that scattering
corrections are negligible \citep{WKC2000}. This category is independent of the morphological
categories, and subsets of each type fulfill the $\Delta z$ criterion. The electronic table lists this
in a separate column for ease of search.

\begin{figure*} 
\includegraphics[width=165.mm]{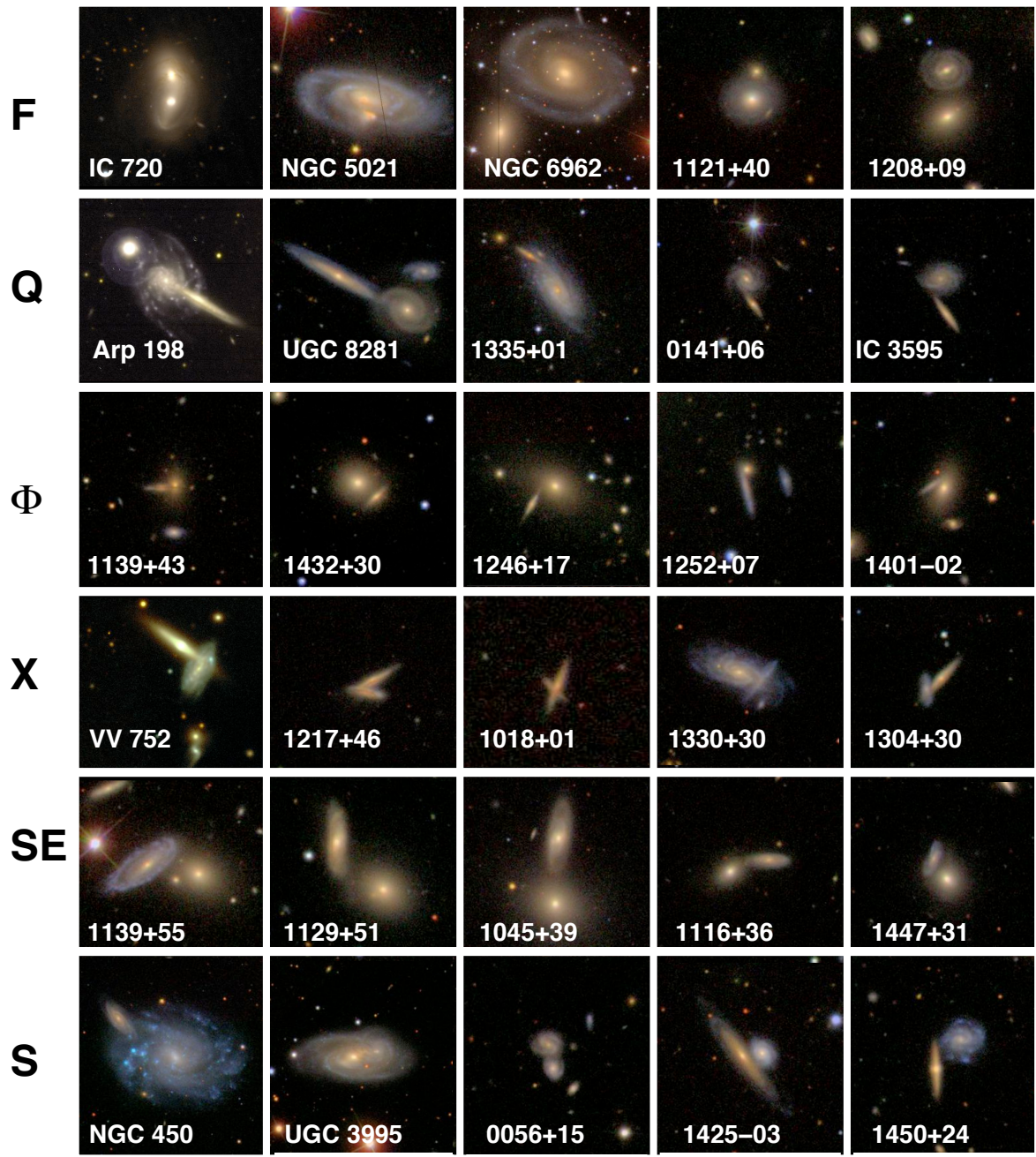} 
\caption{Sample pairs of various geometrical types. Images are from the JPEG files delivered by the SDSS SkyServer  or our WIYN imaging displayed 
with similar intensity mapping. North is at the top; image sizes are 1, 2, or 4 arcminutes square. Pairs are identified by common name or truncated coordinate designation to facilitate finding them in Table \ref{tbl-catalog}. (Continued on next page)} 
\label{fig-montage1} 
\end{figure*} 

\addtocounter{figure}{-1}

\begin{figure*} 
\includegraphics[width=165.mm]{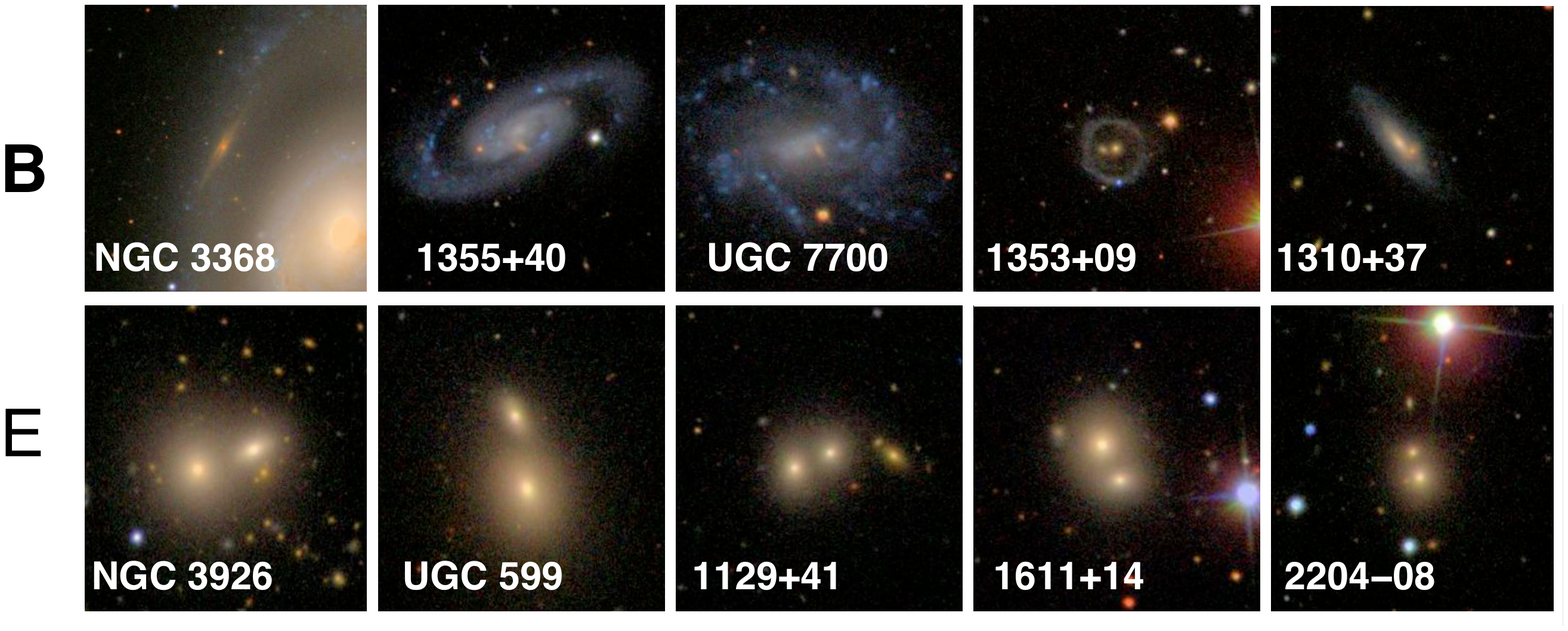} 
\caption{(continued)} 
\end{figure*} 

Among several of these categories we find deep overlaps, pairs in which the background galaxy is seen through the inner part of the foreground system. The reddening is often so strong that they
can be selected from the SDSS $gri$ composites for showing the background
nucleus only at $i$. The nearby type example is NGC 3314 \citep{KW2001b}; this catalog includes several additional examples of central spiral/spiral overlaps, as well as some fairly nearby instances of backlighting extending almost to the foreground nucleus. These objects are important (although we do not know how complete the selection is) because of evidence that the inner parts of some disks have significant optical depth even in the $K$ band, making further IR observations a
promising approach for extinction measurements very deep inside these spirals. Table \ref{tbl-deep} lists such deep overlaps from our catalog which are like
NGC 3314 in the sense of having a background galaxy large and bright enough, projected close enough, for high-surface-brightness backlighting of the inner regions (typically inner 3 kpc) of a foreground spiral which is not nearly edge-on. Six of these fall in our type R, which may also be polar ring systems. Table \ref{tbl-deep} includes
the projected angular separation between galaxy nuclei.

\begin{figure*} 
\includegraphics[width=170.mm,angle=0]{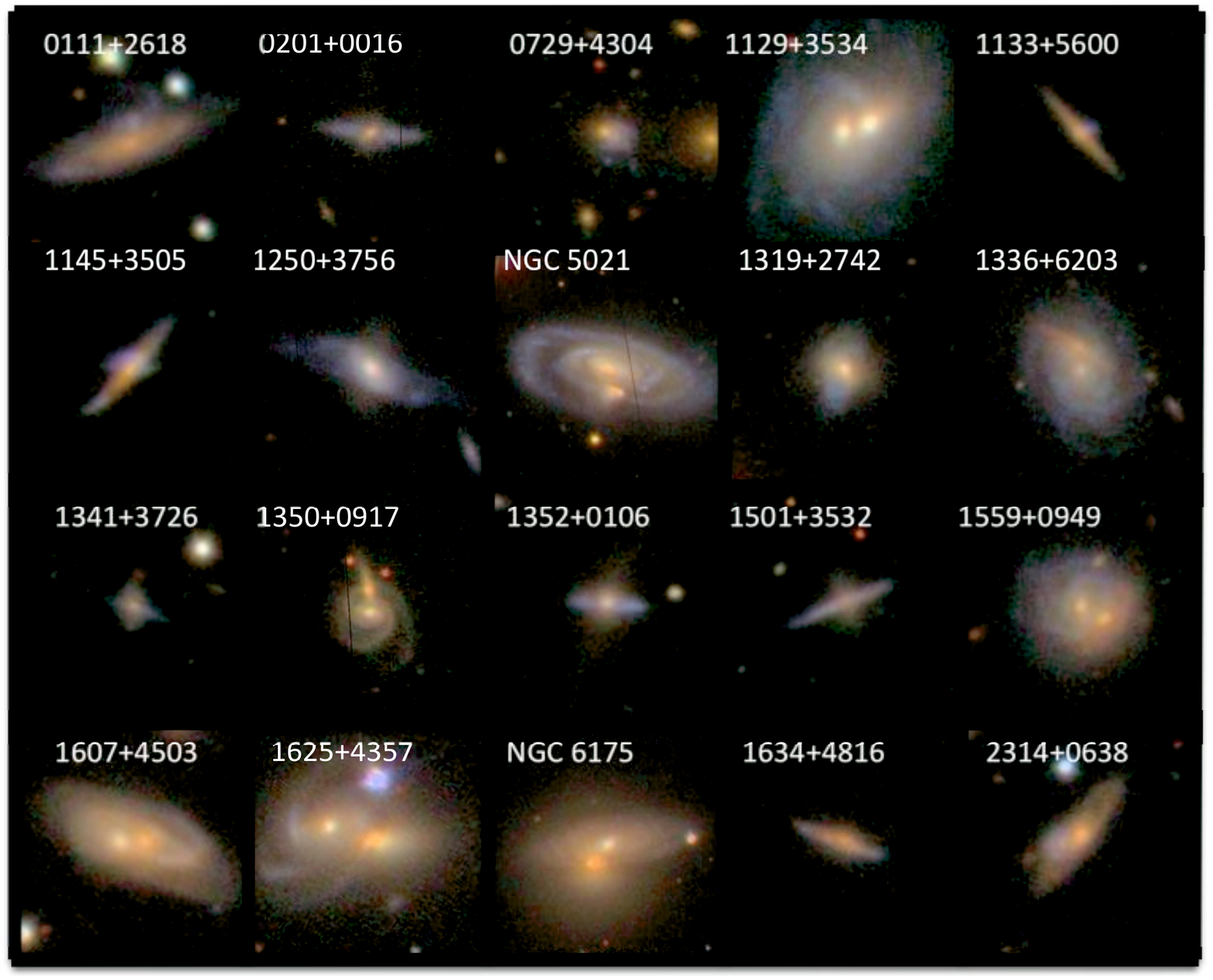} 
\caption{Deep overlaps, similar to NGC 3314 in having the galaxies' nuclei projected well within the area of the foreground disk, so there is a 
bright background source to study the extinction deep within a disk which is not itself edge-on (listed in Table \ref{tbl-deep}). 
Images are $gri$ color composites from the SDSS
Sky Server; all are 50" square except NGC 5021 and 6175 which are 100" square. Six of these fall in catalog category R, and might be similar to
polar-ring systems rather than being two independent galaxies; redshift information would distinguish these. NGC 3314 itself, lying outside
the SDSS region, is not shown here; it appears in Fig. \ref{fig-oldpairs}. } 
\label{fig-deep} 
\end{figure*} 

It is difficult to make an external assessment of the completeness of our catalog in magnitude,
since the SDSS photometric pipeline sometimes fails in apportioning flux between deeply
blended objects. Likewise, the pipeline size measurements, such as Petrosian radius, 
are not reliable for much of the sample.

\subsection{Redshifts}

Since photometric redshifts can be accurate enough, in some applications, to pick out galaxy pairs with matching
redshifts \citep{QW2009}, we have explored the use of the SDSS photometric redshift estimators to distinguish additional
pairs with large redshift differences, with disappointing (although not wholly unexpected) results. For
156 distinct pairs having two spectroscopic redshifts from SDSS DR7, we considered the template-based and
neural-network estimates {\it photoz, photoz2} and their error estimates from the SDSS database. Of these, 33
showed catastrophic failures in both estimators, with error $> 5 \sigma$, traceable to deblending
problems. In these cases, at least one pair member has its Petrosian radius underestimated by
factors $>4$, due to the deblending algorithm assigning only the nucleus to this
object and the additional diffuse flux to the other galaxy or to neither. This generally carries
corresponding photometric errors; a strong correlation between errors for the two estimators indicates that the photometric errors
are mostly responsible. Even for pair members without obvious photometric problems,
the photometric estimates provide almost no information for $z < 0.1$ (where both are biased
upward, by 0.03 for {\it photoz} and 0.04 for {\it photoz2}), and are useful only for $\Delta z > 0.15$ (in which case the apparent sizes of the galaxies
otherwise suggest large differences independently).

\begin{figure} 
\includegraphics[width=64.mm,angle=90]{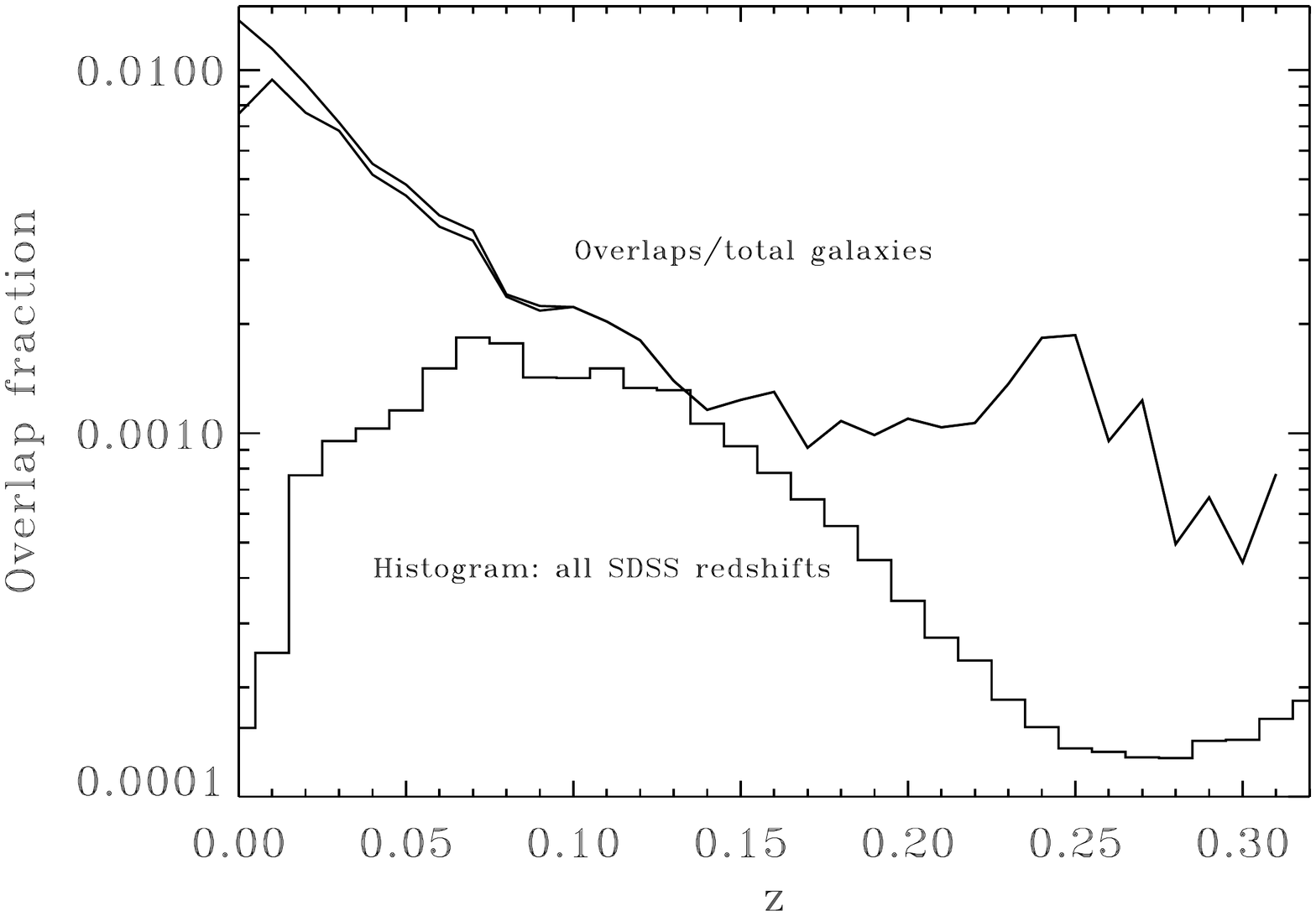} 
\caption{Fraction of SDSS galaxies with measured redshifts $z$ included in this catalog as a function
of $z$. The lower, diverging trace at small $z$ shows the result of omitting type B pairs, where the smaller background galaxy would be missed
at larger $z$. The histogram shows the $z$-distribution of all galaxies with spectroscopic redshifts in the SDSS
(as of DR8), scaled down by a factor $10^{7.5}$ to fit on the same axes. Where both redshifts in a pair
are known, the lower value is used, since the closer galaxy is the one under study. The selection is most complete at low $z$,
exacerbated in this view by the incompleteness of SDSS selection of redshifts for very bright galaxies. A
plateau appears at $z=0.1-0.22$. Above this, the SDSS parent population changes as most galaxies observed were
color-selected to be luminous red galaxies (LRGs).} 
\label{fig-zcomp} 
\end{figure}

Our best assessment of the completeness of sample selection at this point comes from the subset with
spectroscopic redshifts, which we can compare to the entire SDSS to ask what fraction of galaxies
are selected as a function of $z$. Fig. \ref{fig-zcomp} shows the fraction of SDSS galaxies in our
catalog as a function of redshift. We take the lower redshift of a pair when both values are known, since the
foreground galaxy is the one under study. A galaxy is 10 times more likely to be selected at $z=0.01$ as at $z=0.1$, although the
probability stay roughly constant from $z=0.1-0.2$.
We can investigate how much of this effect is due to B-type pairs, with a faint background system; for more distant foreground systems,
there is less volume available for background systems within the effective distance limit for seeing structure in SDSS images.
This effect is significant only for $z < 0.02$; the lower trace in Fig. \ref{fig-zcomp} shows the result of omitting the B-type foreground systems
where the smaller system would be missed at greater distances. This accounts for only a small fraction of the selection change with $z$. A large
part of the effect must be in resolution, as the characteristic structures picking out two galaxies, and distinguishing spirals from
ellipticals,  become so blurred that only
more obvious pairings are found.

\subsection{Magnitude selection}
\begin{figure} 
\includegraphics[width=64.mm,angle=90]{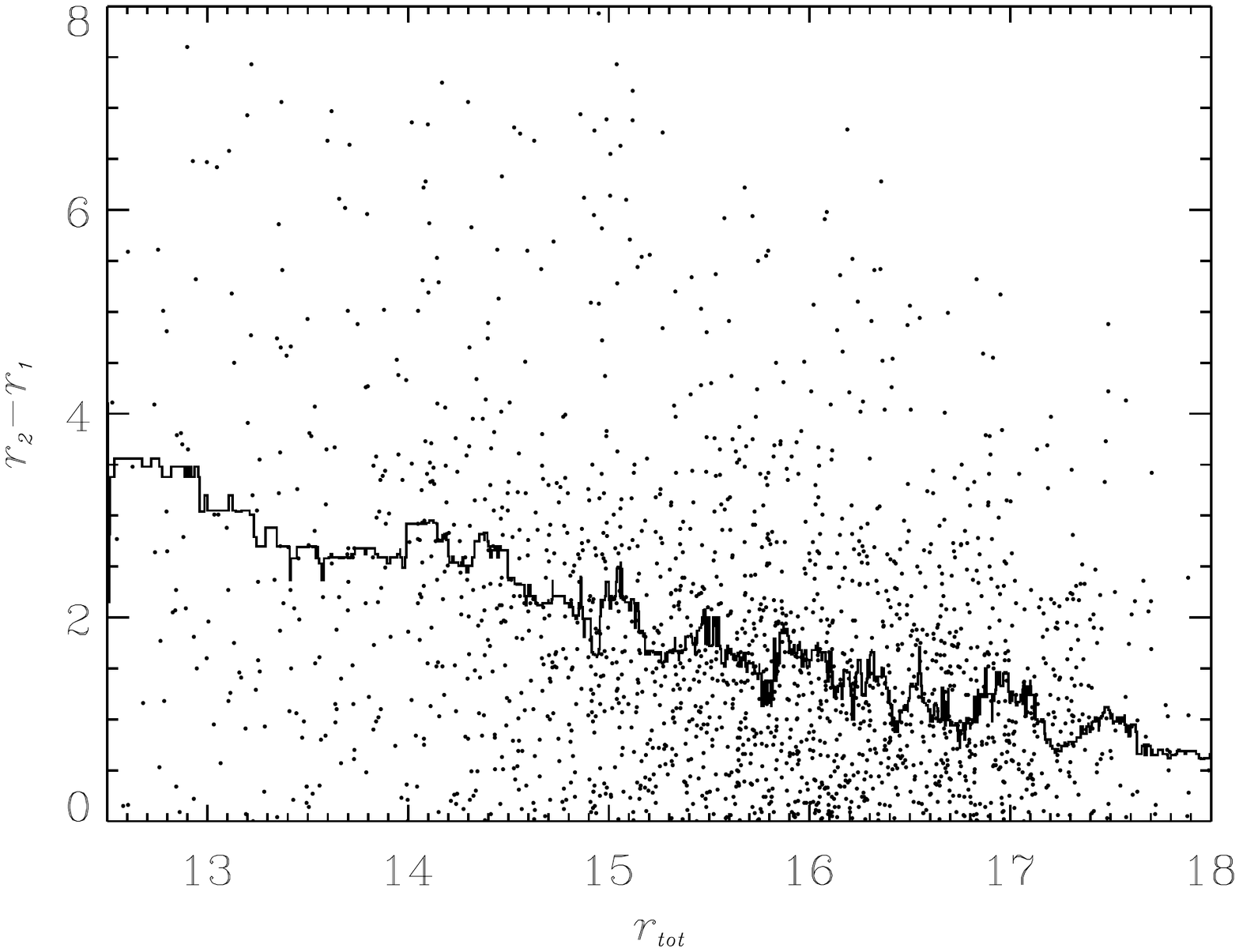} 
\caption{Component magnitude differences $r_2 - r_1$ as a function of integrated magnitude
$R_{tot}$ of sample galaxy pairs, over the well-sampled region $r_{tot}=12.5-18$. The
jagged line is a 51-entry running median, showing a gradual decrease for fainter
pairs in which more nearly equal members are preferentially selected. The range
$r_2 - r_1 < 3.5$ remains well populated over this whole magnitude range.} 
\label{fig-magdiffs} 
\end{figure}

Statistics with pair and component magnitude provide some insight into the sample selection, subject to the caveats above about
pipeline magnitudes for blended images. Following \cite{Darg}, who deal with magnitudes for merging systems, we use the magnitude
$r_{tot}$ of the $r$-band light from both components of each pair, which is more robust to how the light was deblended into constituent
galaxies. The magnitude difference between pair members shows, as expected, a systematic decrease for fainter pairs, albeit with
large scatter, the whole range $r_2 - r_1 < 3.5$ remains well populated for objects brighter than $r_{tot} = 18$ (Fig. \ref{fig-magdiffs}.

\begin{figure} 
\includegraphics[width=64.mm,angle=90]{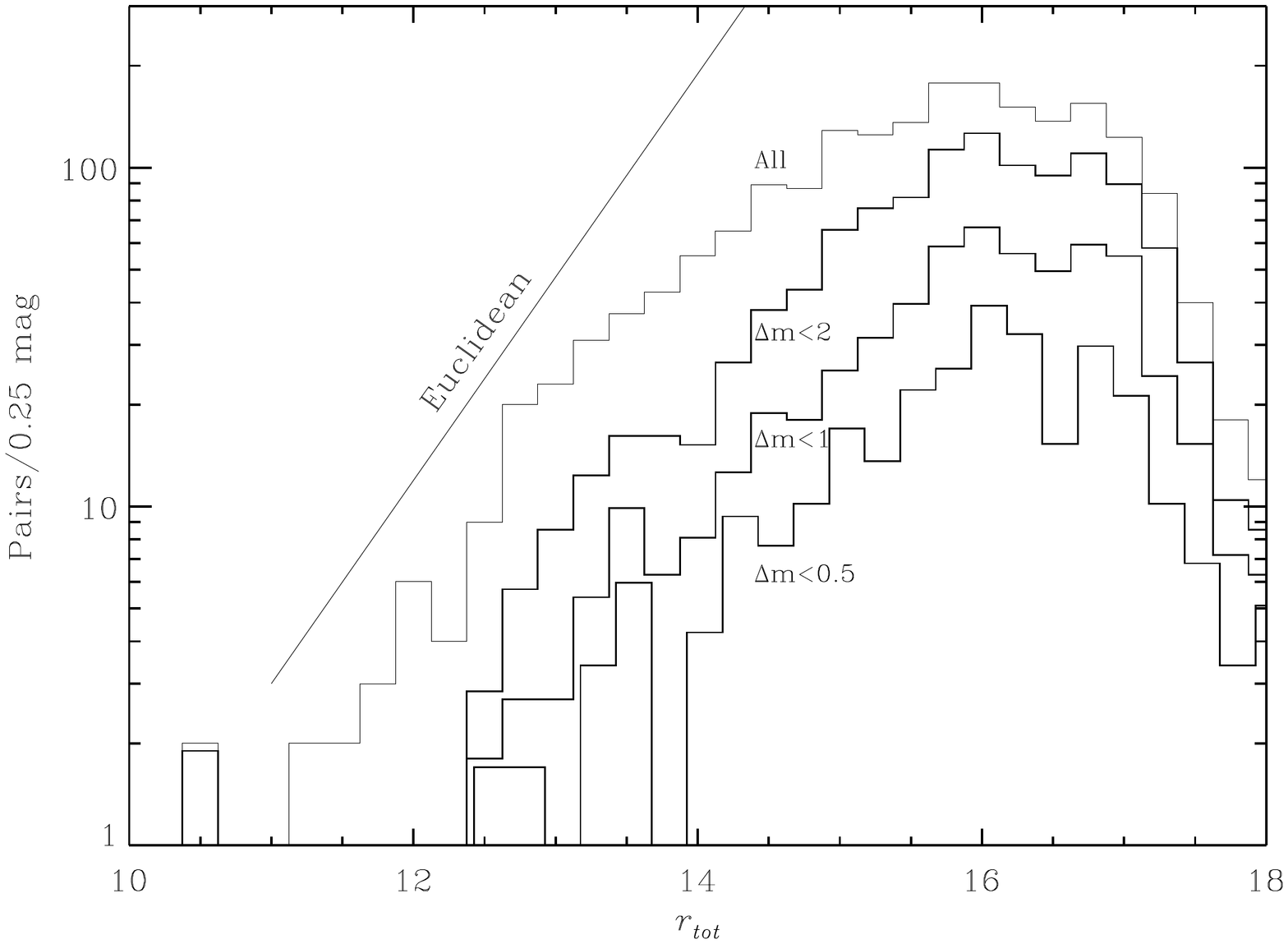} 
\caption{Distribution of sample galaxy pairs with total magnitude $R_{tot}$. The Euclidean
slope is shown for reference in evaluating incompleteness, with the caveat that time-dilation
effects become important in the magnitudes for $z > 0.1$. Subsamples restricted in
component magnitude difference $\Delta m$ are shown, with slight offsets for clarity.} 
\label{fig-magdist} 
\end{figure}

The counts of sample pairs with magnitude follow the Euclidean slope only at the bright end (Fig. \ref{fig-magdist}). When separated into
bins by $\Delta m = r_2-r_1$, the distributions remain broadly similar; incompleteness becomes serious at levels $r_{tot} = 13-14$, reaching 
fainter for more nearly equal pairs. Again, as one might expect, more equal pairs are selected preferentially among fainter systems.

\section{Summary}

From thousands of candidates identified by Galaxy Zoo participants using the Sloan Digital 
Sky Survey, we have produced a catalog of 1990 galaxy pairs with geometry of overlap suitable
for use in the study of dust extinction. These are coded to distinguish various kinds of
overlap useful for different aspects of the problem - spirals partially backlit
by E or S0 galaxies, spiral/spiral pairs, systems with the background galaxy edge-on and
radial to the foreground system, and so on. About 11\% of these have known redshift differences large 
enough to eliminate the possibility of interaction with one another, and thus the possibility of
such an interaction causing a breakdown of the galaxy symmetry needed to analyze the extinction.
Use of these pairs to study extinction complements infrared techniques, giving the high
angular resolution of optical observations and retaining sensitivity to even very cold grain populations.
This sample can also be used to search for additional examples of very extended dust structures, such as found by  \cite{Holwerda2009}. 
Additional uses of this listing might include correlation with supernova searches, yielding complementary reddening information,
use of foreground absorption to distinguish which members of galaxy pairs are in the foreground and hence constrain their orbital
location via redshift differences, and use of ``wrong-way" redshift differences as a probe of peculiar motions.

Further papers will deal with analysis of the images of a larger subset of these galaxies
from optical and ultraviolet images, using these additional data to address the spatial distribution and reddening law of the dust in the optical and ultraviolet ranges.

\acknowledgments

This work would not have been possible without the contributions of citizen scientists as part of the Galaxy Zoo project. We particularly wish to thank the contributors to the ``overlapping galaxies"
forum thread whose candidates underlie the major part of this work. Where supplied, we have used a full name, and have otherwise acknowledged contributions by Galaxy Zoo username in a listing at http://data.galaxyzoo.org/overlaps.html. We are grateful to Jean Tate for a very careful reading of our data compilation
and identification of mismatches and missing data.

W.C. Keel acknowledges support from a Dean's Leadership Board faculty fellowship.
C. J. Lintott acknowledges funding from The Leverhulme Trust 
and the STFC Science in Society Program. K. Schawinski 
was supported by the Henry Skynner Junior Research Fellowship at Balliol College, Oxford and by a NASA Einstein Fellowship at Yale,
and gratefully acknowledges support from Swiss National Science Foundation Grant PP00P2\_138979/1. Galaxy Zoo was made possible by funding from a Jim Gray Research Fund from Microsoft and The Leverhulme Trust. Three galaxy pairs in Abell 3558 were
identified on HST data products presented by Nikolaus Sulzenauer on his ``Hubble Unseen"
web collection at http://quarks.maynau.com/nova/sin\_galerie\_hu.shtml; we thank him as well for identifying the field of one of his gallery images. Claude Cornen caught some transcription errors among our redshift values. We thank Rick Johson for permission to reproduce two of his
images of Arp pairs in Fig. 1. Ian Steer discussed additional uses of the catalog.

Funding for the creation and distribution of the SDSS Archive has been
provided by the Alfred P. Sloan Foundation, the Participating Institutions,
the National Aeronautics and Space Administration, the National Science
Foundation, the U.S. Department of Energy, the Japanese Monbukagakusho,
and the Max Planck Society. The SDSS Web site is http://www.sdss.org/. 
The SDSS is managed by the Astrophysical Research Consortium (ARC) for
the Participating Institutions. The Participating Institutions are The
University of Chicago, Fermilab, the Institute for Advanced Study, the Japan
Participation Group, The Johns Hopkins University, Los Alamos National
Laboratory, the Max-Planck-Institute for Astronomy (MPIA), the
Max-Planck-Institute for Astrophysics (MPA), New Mexico State
University, Princeton University, the United States Naval Observatory, and
the University of Washington.

This research has made use of the NASA/IPAC Extragalactic Database (NED),
which is operated by the Jet Propulsion Laboratory, Caltech, under
contract with the National Aeronautics and Space Administration.
Based in part on observations made with the NASA Galaxy Evolution Explorer. 
GALEX is operated for NASA by the California Institute of Technology under NASA contract NAS5-98034



\appendix

\section{Previously known pairs and pairs outside the SDSS}

\begin{figure*} 
\includegraphics[width=150.mm]{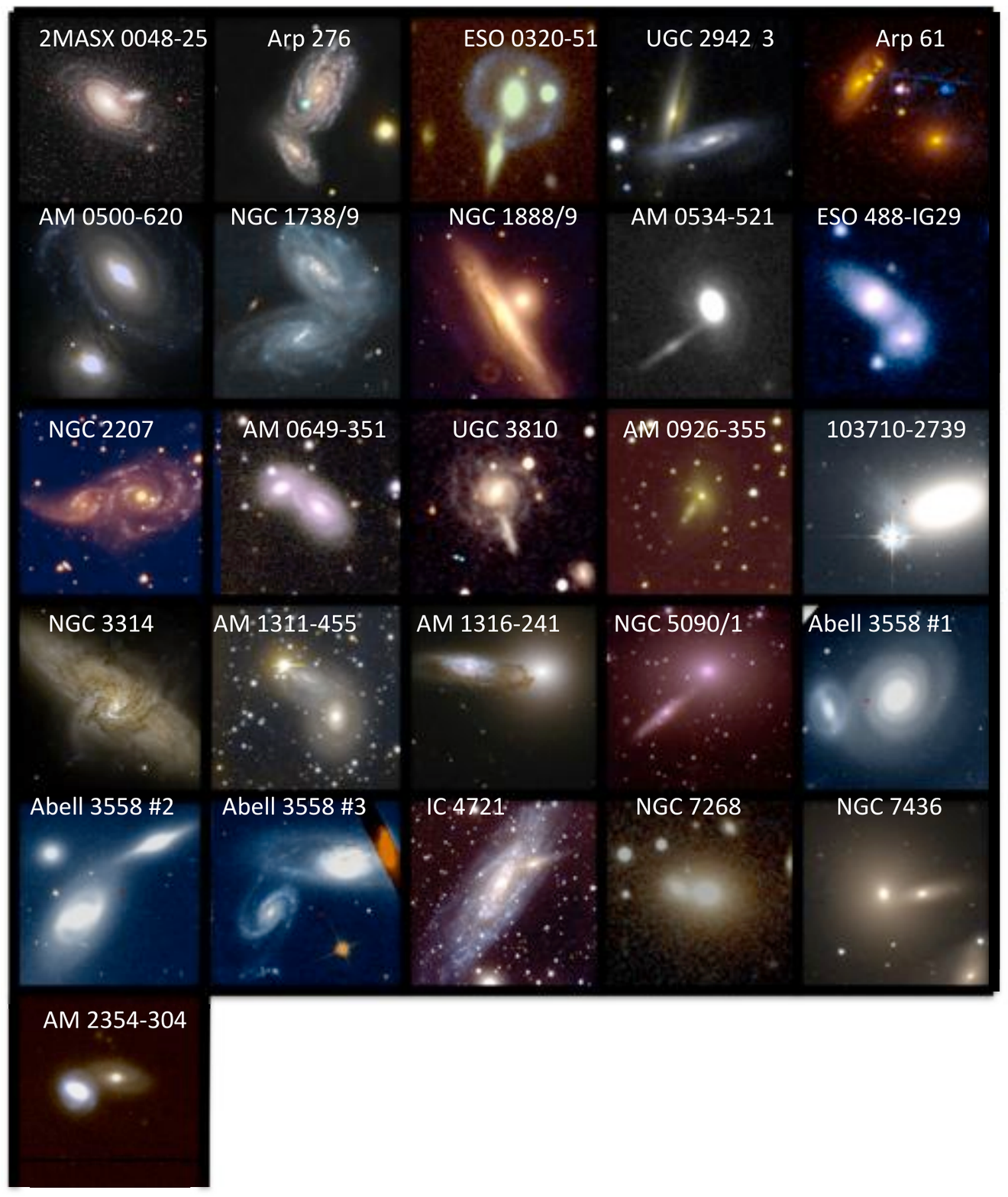} 
\caption{Pairs of overlapping galaxies in the local Universe ($z < 0.07$) outside the SDSS imaging coverage. Where 
possible, the images are similar in construction to the SDSS color composites. Image sources are as given in Table
\ref{tbl-oldpairs}; the image of Arp 276 was obtained by Rick Johnson with his 0.25m telescope. These
may be compared with the SDSS images of both newly found and previously known pairs within the SDSS.} 
\label{fig-oldpairs}
\end{figure*}

For convenience of comparison, we list in Table \ref{tbl-oldpairs} the handful of overlapping
pairs noted in other studies (including 5 recognized serendipitously in HST images, 
at redshifts low enough to be included in the catalog).
Among these, 9 lie within the SDSS sky coverage, all of which were independently found
in the Galaxy Zoo project. Fig. \ref{fig-oldpairs} shows the pairs outside the SDSS DR8 imaging region, collecting color images for ease for comparison.
These are included in the catalog, and may be recognized by their lack of an SDSS ObjID identifier.

\clearpage








 
 \begin{deluxetable}{llccccll}
\tabletypesize{\scriptsize}
\tablecaption{Overlapping Galaxy Pairs}
\tablewidth{0pt}
\tablehead{\colhead{Coordinate name}     & \colhead{SDSS ObjID}  &  \colhead{$r_1$} &  
\colhead{$r_2$} & 
\colhead{$z_1$} & \colhead{$z_2$} & \colhead{Type $\Delta z$} & \colhead{Cross-ID}   }
\startdata
SDSS J000007.03+081645.1& 587743960499880043 & 13.43 & 14.84 & 0.0387 S & & E  & UGC 12890\\
SDSS J000046.97+282407.3 & 758874298530726152 &  13.42 & 14.60 & 0.0272 E & 0.0292 W & F & UGC 12899 \\
SDSS J000058.93+285442.1	& 758874371533308165 & 13.82 & 16.17 & 0.0230 & &	 Q  & UGC 12901\\		
SDSS J000103.67+343911.0	& 758874373141496140	& 14.11 & 19.98 & 0.0423 SE & & 	F\\	
SDSS J000140.21+010531.2	& 587731187814695060	& 15.72 & 21.66 & 0.0611 fg & & $\Phi$\\ 	
SDSS J000253.63+315102.2	& 758874300139962790 & 18.09 & 18.92 & & & Q\\	
SDSS J000320.46+083707.5	& 587743961037078573 & 13.78 & 15.39 & 0.0397 & & 	S & UGC 10\\	
 SDSS J000400.79+160110.4	& 587727223561453822 &  15.95 & 11092 (S) & & F \\
 SDSS J000401.52-111027.3	& 587727177913008261 &  14.75 & 11392 (SW) & & S \\
SDSS J000620.75-105653.7	& 587727225153978471 &  17.73 & 18.32 & 0.1091 NE & & F	 \\
SDSS J000954.49-050116.8	& 587747122131304642 &  16.56 & 21.50 &&  & misc \\
SDSS J001122.29+062321.6	& 587743795144425534 &  12.83 & 15.47 & 0.0201 W &  &	S \\
SDSS J001315.28+000240.7	& 588015509270822939 & 17.03  & 17.78 & 0.0902 SE & 0.0390 NW & Q  $\Delta z$\\
SDSS J001347.12+004612.9	& 587731187279134987 &  16.56 & 21.50 & 0.1556 N & 0.1531 S & E \\
SDSS J001430.96+154907.2	& 587727180601163792 &  16.18 & 16.84 & 0.0806 S&  & S \\				
SDSS J001531.05-004805.6	& 588015508197343379 &  15.15 & 17.64  & 00685 E & 0.1556 W & F $\Delta z$ \\	
SDSS J001725.83-005842.5	& 587731185132109990 &  14.09 & 20.37 & 0.0181 N & & S \\	
SDSS J001747.95-122422.2	& 587747073275003048 &  14.96 & 15.37 & & & F \\
SDSS J002305.73+064506.0	& 587744044784156688 &  14.25 & 17.38 & 0.0494 N &  & S	 \\			
SDSS J002315.77+241310.9	& 587740522933715074 &  17.43& 18.5	& & & X \\

SDSS J005618.88+152531.9 & 587724199351025860  & 16.01 & 16.29  &   0.0709 N  &   0.0771 S  &  S \\
SDSS J005745.03-002509.9 &  588015508738801700 & 13.51 & 14.98 & 0.0408 N & 0.0437 S &  E & UGC 599 \\
SDSS J014145.47+060151.9 & 587744293354668078 & 15.61 & 16.90 &     &    & Q \\
SDSS J101800.73+012116.3 & 587726031700492467 & 17.42 & 30.0 & 0.1265 ctr&  & X \\
SDSS J104524.96+390949.9 & 587735661016318093 & 14.09 & 144.3 & 0.0254 S & 0.0265 N & SE \\
SDSS J104645.67+114911.8 & 588017703996424196&  10.16 &15.73 & 0.0030 SW &   &    S   &  NGC 3368, M96\\
SDSS J105946.49+173912.4 & 587742865816944739 & 14.55 & 17.77 & 0.0295 W  &  0.0294 E  &  Q   &  Arp 198\\
SDSS J111605.47+361410.1 & 587739097520799818 & 15.24 & 16.02 & 0.0776 E & 0.0785 W & SE \\
SDSS J112116.82+402043.4 & 588017719566467080 & 14.31 & 16.49  & 0.0209 S & 0.1147 S &  F $\Delta z$\\
SDSS J112924.32+415219.3 & 588017721177800860 & 15.15 & 15.15 &  &  &  E \\
SDSS J112943.31+511415.7 & 587732134846070798 & 14.08 & 14.29 & 0.0335 E  &  0.0345 W &   SE \\ 
SDSS J113912.22+553957.8 & 587731870706565177&  14.15 & 14.42 & 0.0623 SW &  0.0615 NE &  SE \\
SDSS J113947.32+435031.7 & 588017625615368313 & 16.11 & 17.33 & 0.1344 W &  & $\Phi$ \\
SDSS J114222.33+084611.5 & 587732769978777642 & 13.91 & 13.98 & 0.0216 N & 0.0220 S & F & IC 720\\
SDSS J115128.22+220133.4 & 587742061070057546 & 13.48 & 15.06 & 0.0284 W &   0.0257 E  &  E  &      NGC 3926 \\
SDSS J120802.07+094557.0 & 587732771055337475 & 14.52 & 15.27&  0.0694 S  &        &      F    &    UGC 7114\\
SDSS J121748.55+463454.8 & 588298661962973323 & 17.84 & 20.0 & 0.0658 & & X \\
SDSS J123232.82+635238.3 & 587728676858101872 & 14.35 & 18.3 & 0.0099 N &  & B & UGC 7700 \\
SDSS J123706.39+234712.7 & 587742188833210435 & 15.89 & 16.49 & 0.0570 S & & Q & IC 3595 \\
SDSS J124613.57+171012.8 & 587742774026633226 & 14.69 & 16.90 & 0.0804 NW & & $\Phi$ \\
SDSS J125224.50+071053.3 & 588017724940878045 & 16.68 & 17.01 &  0.0806 N &  & $\Phi$ \\
SDSS J130431.54+303417.7 & 587739721901932619 & 15.69 & 16.98 &  0.0622 E    & &             X  \\
SDSS J131012.03+373125.4 & 587739098604503163 & 15.21 & 16.82 & 0.0429 NE &  &B \\
SDSS J131158.38+444832.1 & 588017627233124523 & 14.75 & 14.75  & & & Q & UGC 8281 \\
SDSS J131206.27+461146.1 & 588298662503972874 & 12.9 & 17.8  & 0.0283 N  && F  & NGC 5021\\
SDSS J133026.37+300144.4 & 587739709015261237 & 14.53 & 17.5 & 0.0373 NE & & X & UGC 8497 \\
SDSS J133558.58+014348.2 & 587726031722184751 & 14.38 & 15.90 & 0.0228 S & & Q \\
SDSS J135316.39+094017.5 & 587736543089000513 & 17.33 & 18.15 & 0.1349 W & & B \\
SDSS J135507.95+401003.4 & 587736585499246631 & 14.45 & 20.06 & 0.0083 NE &&  B & UGC 8841\\
SDSS J140138.45-022558.8 & 587729777442488436 & 15.08 & 17.63 & 0.0505 SW & & $\Phi$ \\
SDSS J142459.13-030401.2 & 587729776371302561& 14.87 & 15.07 & 0.516 SE & 0.0513 NW &      S \\	                                
SDSS J143243.04+301320.8 & 587739380987592822 & 14.52 & 16.16 & 0.0623 (NE) & &$\Phi$ \\
SDSS J143243.90+301329.2 & 587739380987592820 & 14.52 &16.16 & 0.0623 NE   &    &            Phi  \\
SDSS J144230.35+222110.7 & 587739809952890969 & 14.51 & 15.54 & 0.0507 NE & & X & VV 752 \\
SDSS J144750.76+314553.3 & 587739132421799955 & 14.63 & 16.58 & 0.0557 NE &  0.0454 SW &  SE   $\Delta$z &  IC 4508  \\
SDSS J145013.12+241919.0 & 587739720301871290 & 15.26 & 15.38 & 0.0329 NW  &  0.0329 SE &  S  \\
SDSS J161117.22+141531.5 & 587739845394366485 & 14.01 & 14.98 & 0.0329 N &        &       E  \\
SDSS J204719.06+001914.8 & 587731173842026987 & 12.37 & 12.63 & 0.0140 NW & 0.0127 SE & F & NGC 6962 \\
SDSS J220412.01-083836.0 & 587730816822411334 & 15.19 & 16.86 & 0.0640 S &  0.1353 N &  E $\Delta z$ \\
\enddata
\label{tbl-catalog}
\end{deluxetable}

\clearpage

\begin{table}
 \centering
  \caption{Summary of galaxy-pair subsamples}
  \begin{tabular}{@{}llr@{}}
  \hline
  Code & Description & Number \\  \\
 \hline
F  & Face-on spiral and background E/S0  & 369\\
Q  & Background galaxy edge-on and radial &   237\\
$\Phi$ & Foreground disk edge-on and background E/S0 & 156\\
X  & Crossing edge-on disks            &  200\\
SE  & Spiral/elliptical pairs not otherwise listed & 102\\
S   & Two spirals &584\\
B  & Background galaxy has small angular size & 181\\
E   & Two E/S0 galaxies           & 59\\
R   & Could also be polar ring &   6\\
$\Delta z$ & Redshifts indicate not physically associated & 218\\
\hline
\end{tabular}
\label{summary}
\end{table}

\begin{deluxetable}{rlcccclcl}
\tabletypesize{\scriptsize}
\tablecaption{Deeply overlapping galaxy pairs}
\tablewidth{0pt}
\tablehead{\colhead{Coordinate designation} & \colhead{SDSS ObjID}    & 
\colhead{$z_1$} & \colhead{Obj} & \colhead{$z_2$} & \colhead{Obj} & 
 \colhead{Type $\Delta z$} & \colhead{Nuc sep"} 
& \colhead{Cross-ID}}
\startdata
SDSS J011128.81+261816.6 & 758877156282466365 &    & & & &                      S  	&       4.3             \\        
SDSS J020149.25+001646.7 & 587731513147785324 & 0.0434  & & &      &           R  	&    $ <2$      \\                        
SDSS J072900.48+430402.7 & 587738066725896649 & 0.1885  & abs & 0.1827 & em  & SE  &	      2.8 \\                  
              103713.20-274104.0   &                                            & 0.0095  & fg  &  0.0155 & bg &  S     &      2.9  &  NGC 3314\\	                
SDSS J112917.08+353432.6 & 587739305286303747  &  0.0346 &  NW & 0.0345 & SE   &   S  	&      5.8 & NGC 3695\\                      
SDSS J113336.25+560030.9 & 587731889502748827 & 0.0609 & & &     &    X 	 &     1.2 \\                      
SDSS J114512.95+350510.4 & 587739304213938297 &  0.0674 & & &      &     S  	 &     1.0 \\                   
SDSS J125020.69+375656.1 & 587739098602733640 & 0.0351 & ctr & &   &      R   &      $ <1$   &   IC 3828  \\          
SDSS J131206.27+461146.0 & 588298662503972874 & 0.0283 & N  & &       &        F  	  &    9.2 &   NGC 5021 \\                         
SDSS J131940.07+274221.7 & 587741603111632925 &  0.0231 & & &             &     R  	  &   $<1$ \\                    
SDSS J133642.56+620337.3 & 588011219671908454 &  0.0311 & fg & 0.01037 & bg  &  S  $\Delta z$ &	      5.5 \\                    
SDSS J134134.82+372625.7 & 588017977826345079 &  0.0628  & fg & 0.1713 &  bg     &           S  $\Delta z$ &	      1.2   \\                    
SDSS J135031.55+091704.1 & 587736542551802124 & 0.0653 & N  & &      &          Q  &	      4.7 \\                    
SDSS J135239.37+010609.1 & 588848901535629486 & 0.0716 & & &          &        R &  	     $<1$     \\                  
SDSS J150153.48+353239.5 & 587736586042212586 & 0.0496  & & &          &       R  	&    $<1.5$ \\    
SDSS J153322.82+332933.3 & 588017604701257882 & 0.0769 & fg &  &  &   R  &    $<2$" \\
SDSS J155929.99+094900.6 & 587742610274845009 & 0.0731 & fg & 0.1025 &    bg    &      S  $\Delta z$     &    5.2 &  CGCG 079-018\\	            
SDSS J160737.22+450355.2 & 588011101034774582 &  0.0441 & W  &   0.0442   & E  &  S  	  &    5.5   \\                          
SDSS J162557.25+435743.5 & 587729753278578838 &  0.0320& E & & &           F  	&     10.1                \\       
SDSS J162957.52+403750.5 & 587729652348223597 &  0.0300 & NW  &   0.0289 & SE &  F     &     10.2  & NGC 6175	\\                
SDSS J163451.17+481623.7 & 587725994646634705 & & & &                &        R	&     $<1$                         \\
SDSS J231444.84+063821.3 & 587743958884286603 & & & &           &      S  	&      4.4       &               \\
\enddata
\label{tbl-deep}
\end{deluxetable}

\begin{deluxetable}{rcllllllll}
\rotate
\tabletypesize{\scriptsize}
\tablecaption{Previously known overlapping pairs}
\tablewidth{0pt}
\tablehead{\colhead{Coordinate designation}     & \colhead{$z_1$}  &  \colhead{obj} &  
\colhead{$z_2$} & \colhead{obj} & \colhead{Type $\Delta z$} & \colhead{Cross-ID}        &    
\colhead{Image}  & \colhead{References}}
\startdata
\sidehead{Outside SDSS coverage:}
      004821.85-250736.5     &               0.064 & fg  & 0.065 & bg &  S    &    2MASX, in NGC 253 field  &            HST   & Holwerda et al. 2009 \\
      022811.1 +193557          &           0.0138 & N  &  0.0134 & S  &  S      &  NGC 935/IC1801; Arp 276    &          0.25m   \\
      032140.55-513934.2        &           0.058 &  N   &      &     &  Q     &   ESO 0320-51, Fairall 299      &       CTIO1.5 &  WKC \\
      040409.75+220753.7       &          0.0212 & NW  & 0.0212 & SE &  X    &    UGC 2942/3              &             CTIO1.5 &  WKC\\
      043654.7 -021647.0         &         0.0325 & SE    &    &   &    B    &    Arp 61, UGC 3104         & SARA-N         \\     
      050033.90-620350.0        &           0.0281 & S     &  &    &     F     &   AM0500-620, ESO 119-IG27     &        HST  &  WKC, KW2001a\\
      050146.70-180925.4       &        0.0133 & NW  & 0.0130 & SE &  S     &   NGC 1738/9               &            CTIO1.5 &  WKC, D2000\\
      052234.4 -112958           &        0.0081 & SW  & 0.0083 & NE &  SE     &  NGC 1888/9         &                  WIYN  &    D2000\\
      053546.5 -520832          &         0.0147   &     &  &   &     Q     &   AM 0534-521               &           CTIO 4m    &   \cite{Keel85}\\
      054844.25-253335.8       &         0.0370 & SW  & 0.0413 & NE &  S    &    ESO 488-IG29            &             SARA-S &   WKC, D2000\\
      061622.03-212221.6       &            0.0091 & W   & 0.0092 & E  &  S    &    NGC 2207/IC 2163             &        SARA-S &  \cite{Elmegreen}, D2000\\
      065054.41-352056.1        &     0.0345 & NE  & 0.0333 & SW  & S     &   AM 0649-351   &    SARA-S  &  D2000\\
      072151.8 +465038            &       0.0336    &      &  &  &    Q     &   UGC 3810           &                  SARA-N &  \cite{Keel85}\\
      092821.7 -361002             &                       &    &  &  &    $\Phi$   &   AM 0926-355                   &       SARA-S\\
      103710.27-273927.2        &                      &      &  &  &    F     &   in NGC 3314 HST field    &            HST\\
      103713.20-274104.0        &        0.0095 & fg &  0.0155 & bg &  S    &    NGC 3314      &                       HST &  \cite{Keel83}, WKC, KW2001b\\
      131423.6 -460646.3          &      0.0104 & SE  & 0.0103 &  NW &  S    &    AM1311-455      &       CTIO1.5 &  WKC\\
      131932.6 -242914            &      0.0346 & SE  & 0.0320 & NW  & F     &   AM 1316-241    &        HST &  WKC, KW2001a,D1999, WK92\\
      132112.80-434216.4       &           0.0114 & E  &  0.0118 & S  &  SE    &   NGC 5090/1, AM1318-432        &       CTIO1.5  &  D1999\\
      132758.99-312912.4         &                      &    &  &  &     S    &    in Abell 3558   &           HST\\
      132801.02-314458.4         &                       &    &  &  &      Q     &  in Abell 3558      &        HST\\
      132802.02-314650.3        &                       &    &  &  &      S     &  in Abell 3558     &         HST\\
      183424.7 -582948             &      0.0075  & E    &  &  &             Q      &  IC 4721, AM1830-583                & CTIO1.5\\
      222541.40-311202.0       &         0.0294 & B   & 0.0281 & A  &  F  &      NGC 7268      &       SARA-N     &            D2000\\
      225756.90+260900.0       &          0.0245 & B  &  0.0256 & A  &  $\Phi$  &    NGC 7436                       &      WIYN  &   D2000\\
      235727.30-302737.0       &          0.0307 & fg &  0.0614 & bg  & S    $\Delta z$  & AM 2354-304                    &      CTIO   &  WKC, D2000\\
\sidehead{Recovered within SDSS:}\\
SDSS J011530.44-005139.5 &   0.0059 & fg &  0.0381& bg  & S    $\Delta z$ &  NGC 450/UGC 807 & & WKC \cite{Andredakis}\\
SDSS J074409.12+291450.6 &   0.0158 &  W  &  0.0160 & E  &  S    &    UGC 3995   & & WKC, \cite{Marz99}, D2000 \\
SDSS J105945.17+173901.6 &   0.0295 & W  &  0.0294 & E  &  Q     &   Arp 198    & & D2000\\
SDSS J123634.26+111419.9 &    0.0075 & E  &  0.0075 & W  &  S      &  NGC 4567/8  &   &  WKC\\
SDSS J124332.49+113456.6 &  0.0037 & SE &  0.0047 & NW &  F    &    NGC 4647/9  &  &    WKC\\
SDSS J132921.43+372450.4 &   0.0568 & N  &  0.0569 & S  &  S  &      Arp 40, IC 4271  &  &  WKC\\
SDSS J141702.52+363417.7 &   0.0101 & W &   0.0103 & E  &  S  &      NGC 5544/5      &     &  WKC, D1999\\
SDSS J172243.81+620957.8 &  0.0283 & S &   0.0268 & N  &  Q  &      Arp 30, NGC 6365 & & D2000\\
SDSS J225505.54-002454.0  &  0.067 & SE  &  &   &        S  &      MCG-2-58-11& &  D2000\\
\enddata
\label{tbl-oldpairs}
\end{deluxetable}

\
\end{document}